\documentclass[twocolumn,
reprint,
 amsmath,amssymb,
 aps,
]{revtex4-1}
\usepackage{graphicx}
\usepackage{dcolumn}
\usepackage{bm}
\usepackage{color}
\usepackage{amsmath}

\begin{document}

\preprint{APS/123-QED}

\title{Controlled phase separation via density-dependent attractive forces \\}

\author{Gavin Melaugh, Davide Marenduzzo, Alexander Morozov, and Rosalind J. Allen}
\affiliation{%
 School of Physics and Astronomy, University of Edinburgh\\
}%




\date{\today}

\begin{abstract}
Mechanical interactions between biological cells may be mediated by secreted products, making them dependent on the local particle density. Here, we explore the generic physics of density-dependent attractive interactions. We show using Brownian dynamics simulations that density-dependent interactions can produce interesting phase separation behaviour including control of aggregate size during a spinodal decomposition-like separation process. We show that these results are generic using continuum modelling of the appropriate Cahn-Hilliard equation. Our study suggests that density-dependent interactions can provide a generic mechanism for control of aggregate size during phase separation. 
\end{abstract}

\pacs{Valid PACS appear here}
\maketitle



Living systems often exhibit behaviour that is very different from that of non-living matter. Non-equilibrium collective behaviour arising from motility or growth
has inspired a recent surge of interest in the physics of active matter \cite{Ramaswamy2010,Cates2015,Romanczuk2012,Stenhammar,Marchetti,Redner,Schwarz-Linek2012a,Howse,Yeomens}. Although it been less studied among physicists,  the biochemistry of life can also lead to interesting collective behaviour. Living cells can produce a plethora of molecules that can act as signals, modifying the behaviour of neighbouring cells (e.g. quorum sensing signals among bacteria \cite{QSrev1,QSrev2}, which can influence aggregate formation \cite{chemotaxis}), or can directly change the physical interaction between cells (e.g. the production of polymers which can mediate bridging or depletion interactions between bacteria \cite{Flemming2007a,Flemming2010a,Schwarz-Linek2010,Dorken,Wuertz,Mara2003}). 

\begin{figure}[t!]
\begin{center}
\includegraphics[scale=1.0]{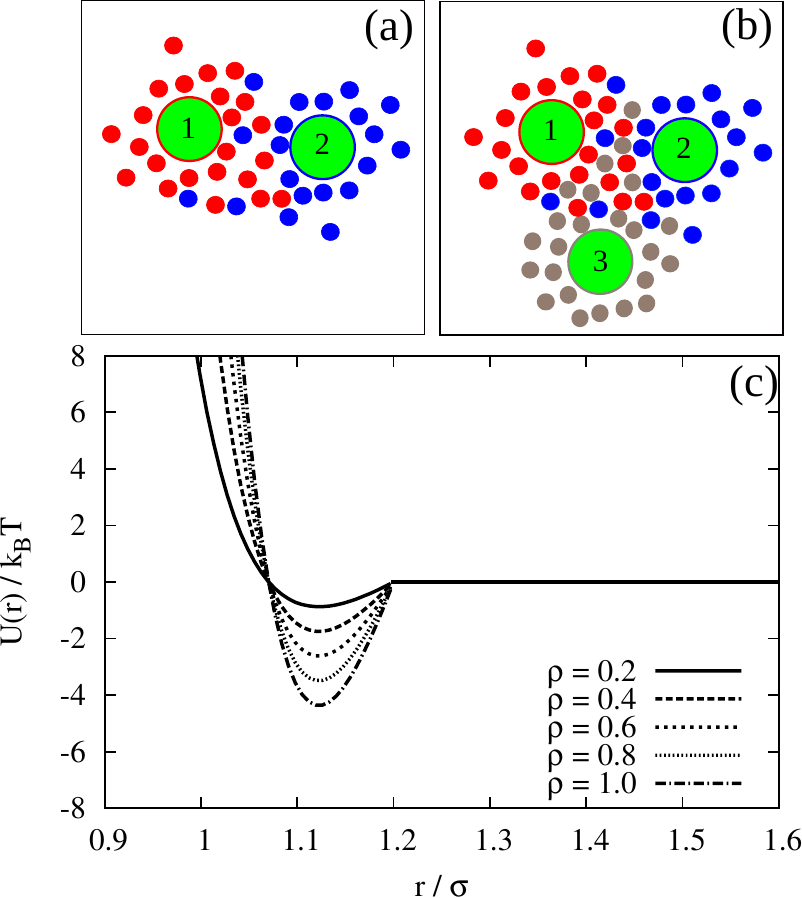}\\
\caption{\label{fig:schematic_interactions} (a) and (b) Schematic illustrating how the attractive interaction between two particles, producing some molecule (e.g., signal or polymer), is affected by the presence of a third particle also producing the molecule.  (a) Particles 1 and 2 produce interaction molecules, shown as smaller discs. (b) An increased number of molecules are present to mediate the interaction between particles 1 and 2 when a third particle 3 is nearby. (c) Density dependence of the cut and shifted Lennard-Jones potential, $U(r)$, for an $\epsilon '$ value of $40 k_BT \sigma^2$. The attractive well-depth increases with increasing $\rho$. At high density, $\rho = 1$, this parameter set gives rise to a potential minimum of $\sim 4 k_BT$. The local density $\rho$ has units of $\sigma^{-2}$.}
\end{center}
\end{figure}

The interactions mediated by these secreted molecules need not be pair-wise additive.   Figures \ref{fig:schematic_interactions}(a) and (b) illustrate a scenario in which cells produce a diffusible molecule that alters inter-cell interactions: the concentration of this molecule between two neighbouring cells, and hence the interaction strength, is increased when a third cell, also producing the product, is in the vicinity. Thus, the effective interaction between two cells is expected to depend on the local cell density. Density-dependent interactions can also arise in non-living systems where particles overlap, such as concentrated polymer solutions, but these are often addressed by minor adjustment of pairwise interaction models to extend them into the semidilute regime, or by using polymer-specific models such as reptation \cite{Bolhuis,Pierleoni,Likos,Jusufi}. Few studies have explicitly addressed the generic physics of density-dependent interactions. In this Letter, inspired by density-dependent interactions between living cells, we formulate a simple model that we use to investigate the basic physics of a system with density-dependent, attractive interactions. We find using both particle-based Brownian Dynamics (BD) simulations and continuum modelling of the Cahn-Hilliard equation that  attractive density-dependent interactions can lead to interesting new physics, in particular, control of aggregate size during phase separation.

We begin by considering a two dimensional system of Brownian particles interacting via  an attractive potential which depends on the local particle density. In particular, we use the modified Lennard-Jones potential illustrated in Figure \ref{fig:schematic_interactions}(c):
\begin{equation}\label{eq:lj}
U(r,\rho) = 4\epsilon(\rho)\left[(\sigma/r)^{12} - (1/r)^{6} - U_{\rm c}\right],
\end{equation}
for $r < r_{\rm c}$, and $U=0$, for $r \ge r_{\rm c}$, where $r$ is the inter-particle pair separation. Here, $\rho$ is the local particle density with units of inverse area ($\sigma^{-2}$), $\sigma$ is the particle diameter and $U_{\rm c} = (\sigma/r_c)^{12} - (\sigma/r_c)^{6}$ ensures that $U=0$ at the cut-off distance $r_{\rm c}=1.2 \sigma$. Although this is, formally, a pair-wise potential, we take many-body effects into account by making the interaction strength density-dependent.
For simplicity, we assume that this dependence, $\epsilon(\rho)$, takes the linear form
\begin{equation}
\epsilon = \rho ({\bf x}) \epsilon', 
\label{eq:liner_dep}
\end{equation}
where $\epsilon'$, which has units of energy times area, is a parameter that determines the sensitivity of the attraction to particle density. Here we show that this density-dependent potential leads to interesting phase separation behaviour.  Specifically, in particular regions of the phase diagram, the system separates via a spinodal decomposition-like process into condensed phase aggregates of rather uniform size, immersed in a low-density ``gas-like'' phase. This behaviour is quite different to that of Brownian particles with conventional pair-wise attractive interactions.

To investigate the behaviour of a system of particles with such density-dependent cohesion, we performed two-dimensional BD simulations of $N$ monodisperse discs interacting via Eqs. (\ref{eq:lj}) and (\ref{eq:liner_dep}). The local density, $\rho$, at each particle coordinate, ${\bf x}$, was computed on a grid at each time-step and the corresponding value of $\epsilon$ was assigned to that particle via Eq. (\ref{eq:liner_dep}) (see the Supplemental Material \cite{SI} for details). To compute interactions between pairs of particles with different local densities, and hence different values of $\epsilon$, we used the Lorentz-Berthelot rule  $\epsilon = \sqrt{\epsilon_1 \epsilon_2}$, where $\epsilon_1$ and $\epsilon_2$ were computed according to Eq. (\ref{eq:liner_dep}). We compare our results to those of equivalent simulations with a standard, density-independent, Lennard-Jones potential (no $\rho$ dependence in Eq. \ref{eq:lj}). In this study, our particles are assumed not to be self-propelled (i.e. they are non-motile), and the total system density was conserved throughout the simulations.


The position, ${\bf x}_i$ of an individual particle, $i$, evolves in our simulations via numerical integration of the over-damped Langevin equation  
\begin{equation}
\frac{{\rm d} {\bf x}_i}{{\rm d} t} = \beta D {\bf F}_i + \sqrt{2D} {\bf \eta}_i(t), 
\label{eq:langevin}
\end{equation}
where $D$ is the diffusion coefficient, $\beta = 1/k_BT$, ${\bf F}_i  = -\nabla \sum_{j \ne i}^N U_{ij}$ is the force on particle $i$ resulting from interactions with its $N-1$ neighbours, and ${\bf \eta}_i(t)$ is a unit variance white noise variable with $\langle \eta_i(t) \rangle=0$ and $\langle \eta_{i;\alpha}(t) \eta_{i;\beta}(t') \rangle = \delta_{\alpha,\beta}\delta(t-t')$, with $\alpha,\beta = x,y$
\begin{figure*}[t!]
\begin{center}
\includegraphics[scale=1]{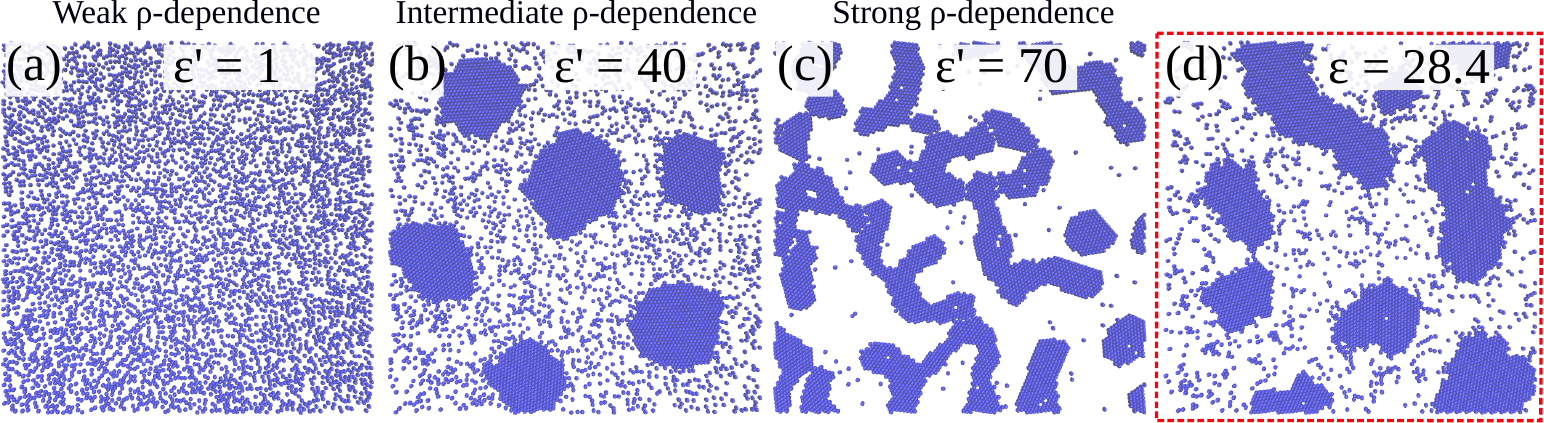}
\caption{\label{fig:two} Simulation snapshots of systems of Brownian particles interacting via a density-dependent potential and a density-independent potential (red box) at $\theta=0.29$. Snapshots show simulation configurations at times $t_{run}=1250\tau$. $\epsilon'$ has units of $k_BT \sigma^2$ and $\epsilon$ has units of $k_BT$.}
\end{center}
\end{figure*}

To implement our simulations, we non-dimensionalized Eq. \ref{eq:langevin} using $\sigma$, $k_BT$, and $\tau=\frac{\sigma^2}{D}$ as the basic units of length, energy, and time respectively (see the Supplemental Material \cite{SI} for details). Simulations were performed in a square box of length  
$115 \sigma$ with periodic boundary conditions, using the Euler method of numerical integration with a time-step of $2.5 \times 10^{-5} \tau$. We performed a systematic study of the effect of the interaction strength parameter, $\epsilon'$, at three different area fractions, $\theta$, determined by the number of particles: $\theta=0.21$ ($N=3600$), $\theta=0.29$ ($N=4900$), and $\theta=0.38$ ($N=6400$). We observed non-trivial phase separation behaviour that is generic to all three systems (see the Supplemental Material \cite{SI} for details); here we present the results for the intermediate area fraction $\theta=0.29$.  

After an equilibration period without inter-particle attractions, simulations were run for times, $t_{run}=1250\tau$, which is the approximate time required for the fraction of particles in the condensed phase to reach a steady state. After time $t_{run}$, coarsening and coalescence changes the distribution of aggregate sizes (ultimately leading to the formation of one large phase-separated domain), but the aggregated fraction remains constant (see the Supplemental Material \cite{SI} for details). The time-scales used here, although not long enough to observe fully equilibrated states, are long enough to capture the rich early aggregation behaviour resulting from density-dependent cohesion. 

Figure \ref{fig:two} shows configurations of the system for $\theta=0.29$ and for various values of $\epsilon'$, after simulation time $t_{run}=1250\tau$. For $\epsilon' = 1$ (weak density-dependent attraction), the system is in a ``gas-like'' phase (Fig. \ref{fig:two}(a)) with ordering, characterised by the structure factor $S(q)$, consistent with that of a simple colloidal dispersion or hard sphere fluid (see the Supplemental Material \cite{SI} for details). For $\epsilon' = 70$ (strong density-dependent attraction), elongated and interconnected aggregates give rise to a ``gel-like'' state (Fig. \ref{fig:two}(c)).


For $\epsilon '=40$ (intermediate density-dependent attraction), we obtain interesting behaviour arising from the density-dependent attraction. It is  clear that the system is undergoing phase separation into condensed and non-condensed phases (Fig. \ref{fig:two}(c)), with the emergence of the former being confirmed by the presence of well-defined peaks in $S(q)$ (see the Supplemental Material \cite{SI} for details). 
Strikingly, we see the presence of condensed phase aggregates immersed in a dilute ``gas''. Phase separation happens in the density-dependent system via the rapid and widespread formation of aggregates that emerge with no apparent barrier to formation (see the Supplemental Material \cite{SI} for movie). Although such behaviour is consistent with spinodal decomposition, phase separation in this regime ($\epsilon '=40$) proceeds without the formation of the elongated structures typically associated with the aggregates formed in the initial stages of spinodal decomposition.


\begin{figure}[t!]
\begin{center}
\includegraphics[scale=1]{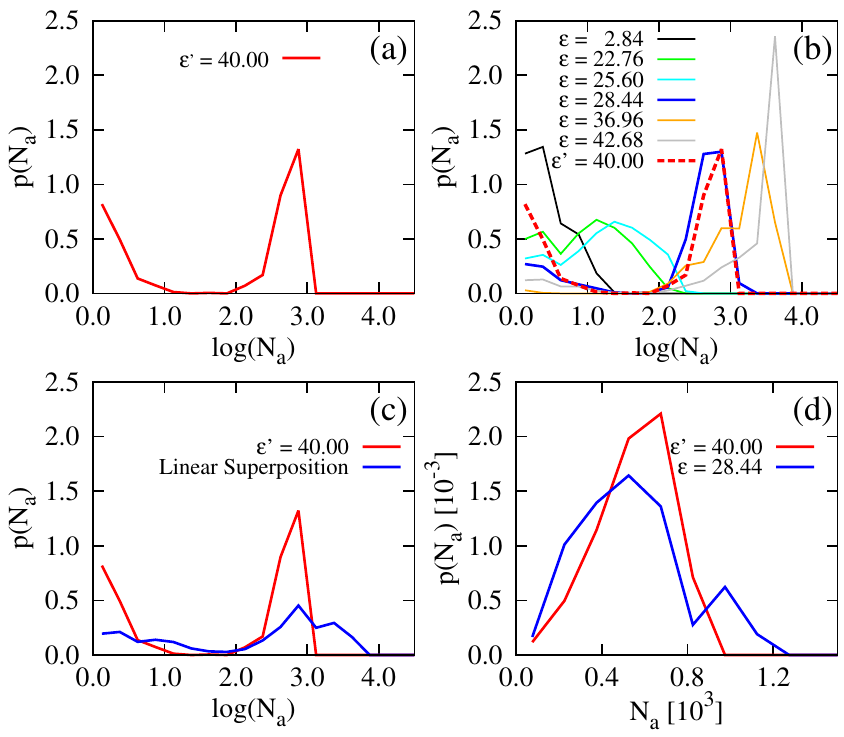} 
\caption{\label{fig:three} Aggregate size distributions, computed at $t_{run}=1250\tau$ for density-dependent and -independent systems at $\theta=0.29$. The quantity $p(N_a$) denotes the  probability that a particle belongs to an aggregate containing $N_a$ particles. In plots (a) to (c), the distributions are plotted vs the logarithm of $N_a$. (a) $p(N_a)$ for density-dependent system with $\epsilon'=40$. (b) $p(N_a)$ for a representative sample of density-independent systems simulated at the $\epsilon$ values explored in a density-dependent simulation with $\epsilon ' = 40$. For comparison, the red curve shows the corresponding distribution for $\epsilon '=40$.
(c) $p(N_a)$ for a weighted linear superposition (blue curve) of the distributions computed in the density-independent simulations of Fig. \ref{fig:three} (b) (see the Supplemental Material \cite{SI} for details).
(d) Aggregate size distribution, including only aggregates of size $N_a>10$ particles, for $\epsilon = 28.44$ and $\epsilon ' = 40$.
All distributions were generated from 10 final configurations at $t_{run}=1250 \tau$ from 10 replicate simulations. $\epsilon'$ has units of $k_BT \sigma^2$ and $\epsilon$ has units of $k_BT$.}
\end{center}
\end{figure}
To quantify the clustering behaviour in Fig. \ref{fig:two}(b), we computed the aggregate size distributions for the $\epsilon'=40$ density-dependent system from 10 configurations at time $t_{run}$ (Fig. \ref{fig:three}(a)), generated in 10 replicate simulations. The distributions shown in Fig. \ref{fig:three} are normalised to provide a measure of the probability that a particle belongs to an aggregate of a given size. The distribution in Fig. \ref{fig:three} (a) is bimodal, with two peaks corresponding to aggregates of $>300$ particles and individual particles (clusters of $<2$ particles). 

The rapid formation of aggregates of rather uniform size, and the bimodal aggregate size distribution appear to result from the density-dependent attractive potential, since these features are not usual in systems with density-independent attractive interactions. To test this idea, we attempted to reproduce these behaviours in an equivalent system with density-independent interactions. To this end, we carried out density-independent simulations with Lennard-Jones interaction strength $\epsilon$ corresponding to the values of $\epsilon$ sampled (via Eq. \ref{eq:liner_dep}) in the final configurations of our density-dependent simulations. For example, in our density-dependent simulations for $\theta=0.29$ and $\epsilon'=40$, the attractive strength $\epsilon$, obtained from the local particle densities via  Eq. \ref{eq:liner_dep}, samples a range from $2.84$ to $42.68 k_BT$.  We therefore chose $\epsilon$ values in this range for our density-independent simulations (see the Supplemental Material \cite{SI} for details).

Figure \ref{fig:two}(d) shows a representative configuration from a density-independent simulation with $\epsilon=28.4k_BT$, undergoing phase separation. Comparing with the corresponding density-dependent simulation (Figure \ref{fig:two}(b)), it is evident that the aggregates emerging from the density-independent system 
are more varied in size and shape. Furthermore, the non-condensed phase contains small clusters but very few ``single'' particles (compare blue and red curves Fig. \ref{fig:three}(b)).


Figure \ref{fig:three}(b) shows the distribution of aggregate sizes from a representative sample of the density-independent simulations with $\epsilon$ values in the range $\epsilon=2.84\to 42.68$ (a complete set of distributions is shown in the Supplemental Material \cite{SI}). None of the distributions that result from the density-independent simulations are bimodal. In these simulations, the system forms either large aggregates, with a broad size distribution  (for $\epsilon > 25.6$), or  predominantly small aggregates, also with a broad size distribution (for $\epsilon \le 25.6$); note here that the peaks located in the range $1<\log(A)< 2.0$ (green and cyan curves) correspond to transient aggregates, i.e., aggregate ``seeds'' which dissolve soon after formation.
None of our density-independent simulations achieved coexistence of large aggregates, narrowly distributed in size, in a ``sea'' of single particles. We also constructed a weighted linear superposition of all the aggregate size distributions generated from the density-independent simulations (see the Supplemental Material \cite{SI} for details); this also fails to match the aggregate size distribution produced by the density-dependent simulations (Figure {\ref{fig:three}(c)). From Figs. \ref{fig:two} and \ref{fig:three}, it is evident that density-dependent cohesion provides a means by which condensed phase aggregates of rather uniform size can coexist with greater numbers of unaggregated particles.



Focusing on the properties of the condensed phase aggregates, in Fig. \ref{fig:three}(d) we plot the aggregate size distribution (on a linear scale) for aggregates of size $N_a>10$ only; the distribution from the $\epsilon=28.44$ simulations is shown as a representative system for the density-dependent simulations as this gives the closest match to the density-dependent case in Fig. \ref{fig:three}(a) (see also Fig. \ref{fig:three}(b)). The size distribution is indeed narrower for the density-dependent case, suggesting that this mechanism of aggregation offers a degree of control in aggregate size.

\begin{figure}[t!]
\begin{center}
\includegraphics[]{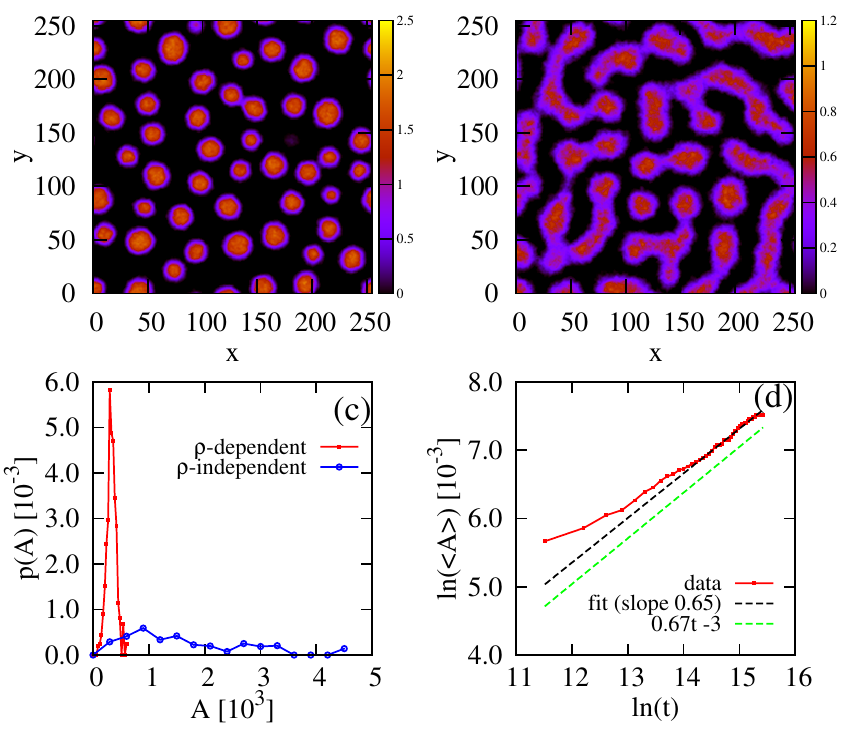}
\caption{\label{fig:four} Continuum model results. All dimensions are in simulation units (SU). Simulation snapshots at $t_{run} = 2\times 10^5$ SU for systems of size $255\times 255$ SU undergoing phase separation interacting with (a) and without (b) density-dependent cohesion. The colour bar is a measure of the density. (c) Distribution of aggregates sizes for the density-dependent and -independent systems were computed at $t_{run} = 2\times 10^5$ SU from simulations of larger system size ($511\times 511$ SU). The probability density provides a measure by which a pixel belongs to an aggregate of size $A$. (d) Log-log lot of the time evolution of the mean aggregate size $\langle A \rangle$ in the density-dependent system again computed from a simulation of size $511 \times 511$ SU. Parameter values for density-independent model (Eq. \ref{eq:non_dens_dep}) are $a=-0.2$, $b=0.2$, $\kappa =0.1$, $M=0.01$, $\xi_0=0.2$, $\Lambda=0.1$, and for density-dependent model (Eq. \ref{eq:dens_dep}) are $a'=0.05$, $b'=-0.5$, $c=0.25$, $\kappa =0.1$, $M=0.01$, $\xi_0=0.3$, $\Lambda=0.1$.}
\end{center}
\end{figure}

The key features of the non-trivial phase separation behaviour emerging from our BD simulations can be captured using a 2-dimensional continuum model for the evolution of the density field $\xi$. We start by noting that density-independent gas-liquid phase separation can be represented with a continuum model using a Landau-Ginzburg free energy density \cite{Chaikin1995} of the form
\begin{equation}\label{eq:non_dens_dep}
f(\xi({\bf x})) =\frac{a}{2}\xi^2+\frac{b}{3}\xi^3+\frac{\kappa}{2}\left(\nabla \xi\right)^2, 
\end{equation}
where $\xi$, the order parameter, is a measure of the local density which, in order to be physically realistic, must satisfy the condition $\xi \ge 0$ (see the Supplemental Material \cite{SI} for details). 
The $\xi^2$ term encapsulates an effective two-body interaction, in which the value of the critical parameter (the parameter that changes sign), $a=\psi-\upsilon$, determines the onset of ordering in the system via a trade-off between entropy ($\psi$) and attractive interactions ($\upsilon$). The second term ($\sim \xi^3$) is the lowest higher order term required to ensure that the equilibrium state has a bounded value of $\xi$, and thus must have a positive coefficient $b>0$. The square gradient term imposes a free energy cost for any non-uniformity in $\xi$, and thus $\kappa$, which is related to the surface tension, must be positive.  
 
Density-dependent interactions can be captured in this Landau free energy formalism by setting the attraction parameter $\upsilon = \upsilon' \xi$ and substituting this into $a$ in Eq. \ref{eq:non_dens_dep}, to give
\begin{equation}\label{eq:dens_dep}
f(\xi({\bf x}))=\frac{a'}{2}\xi^2+\frac{b'}{3}\xi^3 + \frac{c}{4}\xi^4 +\frac{\kappa}{2}\left(\nabla \xi\right)^2,
\end{equation}
where $a'=\psi>0$, $b' = b-\frac{3}{2}\upsilon'$, and the addition of the quartic term with $c>0$ is now required to stabilise the free energy. This form of the Landau free energy is expected to show different physics from the density-independent form (Eq. \ref{eq:non_dens_dep}) because the critical parameter, which changes sign, is now $b'$, the coefficient of the cubic term.

We are interested in situations where the overall density is conserved, e.g., cells that do not proliferate. Therefore the spatial integral of the order parameter obeys $\int_A \xi {\rm d}{\bf x} = \xi_0 A$, where $\xi_0$ is the overall system density and $A$ is the total area, and so the dynamics of $\xi$ can be modelled using the Cahn-Hilliard equation \cite{Cahn-Hilliard}
\begin{equation}\label{eq:cahn_hilliard}
\frac{\partial \xi({\bf x},t)}{\partial t} =   M \nabla^2 \left( \frac{\delta }{\delta \xi ({\bf x})}\int_A f(\xi({\bf x})) {\rm d}{\bf x} \right)+ \nabla \cdot{\bf J}_r,
\end{equation}
where $M$ is the mobility, the term in brackets is the chemical potential, and ${\bf J}_r$ is a random flux which is spatially and temporally uncorrelated, i.e., $\langle {\bf J}_r({\bf x},t) \rangle=0$ and $\langle {\bf J}_{r;\alpha}({\bf x},t) \cdot {\bf J}_{r;\beta}({\bf x},t') \rangle = \Lambda \xi \delta_{\alpha,\beta} \delta({\bf x}-{\bf x}')\delta(t-t')$, with $\alpha,\beta = x,y$. For simplicity, both $M$ ($0.01$) and the random noise strength $\Lambda$ ($0.1$) are kept constant.  
Equation \ref{eq:cahn_hilliard} was solved on an $L \times L$ grid ($128\le L \le 1024$) by means of standard finite difference simulations, with periodic boundary conditions. Both the density-dependent and -independent systems were initialised with a uniform density $\xi=\xi_0$ and quenched into the spinodal regions of the respective phase diagrams, which were generated using the common tangent construction for their respective homogeneous free energy densities (see the Supplemental Material \cite{SI} for details).   

The continuum model, being a macroscopic approach, cannot capture the single particle resolution of our particle based simulations, but it does provides more insight into the behaviour of the condensed phase aggregates. 
Figures 
show representative snapshots within the phase separating regimes in the density-dependent and density-independent systems respectively at time $t_{run} = 2\times 10^4 \tau$, where $\tau$ is now in simulation units (see SI for mapping of simulation units to real units). The condensed phase aggregates here bear a striking resemblance to those in the corresponding phase separating regimes in our particle based simulations (Figs. \ref{fig:two}(b) and (d)). In particular, we see the formation of more rounded aggregates when compared to the elongated structures of the density-independent system. The aggregate size distributions in Fig. \ref{fig:four}(c) confirm that the density-dependent model gives rise to a narrower distribution of aggregate sizes than the density-dependent model.  

As was the case in our particle-based simulations, phase separation resulting via density-dependent cohesion in our continuum model is also initialised from system-wide aggregation events that appear to lack energetic barriers, consistent with spinodal decomposition; but again with the peculiarity that the aggregates are more similar in shape to those that would typically arise from nucleation events. These aggregates continue to grow via coarsening and coalescence such that the time evolution of the mean aggregate size, $\langle N_a \rangle$,  scales as $t^{0.65}$ (Fig. \ref{fig:four}(d)). This is commensurate with classical models of phase separation in diffusive systems with negligible hydrodynamic interactions where typically one finds $\langle N_a \rangle \sim t^{0.67}$ \cite{Chaikin1995}. 

In this Letter, inspired by density-dependent interactions in biological systems, we have introduced a model of density-dependent cohesion in both particle-based and continuum simulations.
Our results show that such interactions give rise to phase separation behaviour that differs fundamentally from that shown by systems with density-independent interactions. In particular, density-dependent cohesion leads to phase separating regimes in which the aggregates sizes are more narrowly distributed and in which aggregates coexist with non-aggregated particles, with a bimodal aggregate size distribution.

Living organisms often interact through secreted molecules, some of which can affect the physics of inter-cellular interactions. Our results suggest that the density-dependent  nature of such interactions could provide a way for micro-organisms such as bacteria to control their aggregation behaviour and form aggregates of well-defined size and shape, with implications for surface colonisation by biofilms \cite{Kragh2016a,Melaugh2016a}, and bacterial infections \cite{Bjarnsholt2009a}. Our results also suggest a mechanism for control of self-assembly processes more generally for systems in and out of equilibrium.


G.M thanks Joakim Stenhammar and Kasper Kragh for helpful discussions in the early stages of this project and gratefully acknowledges financial support from both the EPSRC (Grant No. EP/J007404) and the HFSP (Grant No. RGY0081/2012). R.A was supported by a Royal Society University Research Fellowship and by the ERC under grant 682237 ``EVOSTRUC''. D.M. was supported by the ERC grant 648050 ``THREEDCELLPHYSICS''.


\bibliography{manuscript}

\end{document}



\title{SUPPLEMENTARY INFORMATION \\
Controlled phase separation via density-dependent attractive forces 
}

\author{Gavin Melaugh, Davide Marenduzzo, Alexander Morozov, and Rosalind J. Allen}
\affiliation{%
 School of Physics and Astronomy, University of Edinburgh\\
}%




\date{\today}


\pacs{Valid PACS appear here}
\maketitle


\section{\label{sec:density_grid}Computing the density and the corresponding $\epsilon$}
\begin{figure}[t!]
\begin{center}
\includegraphics[scale=1.0]{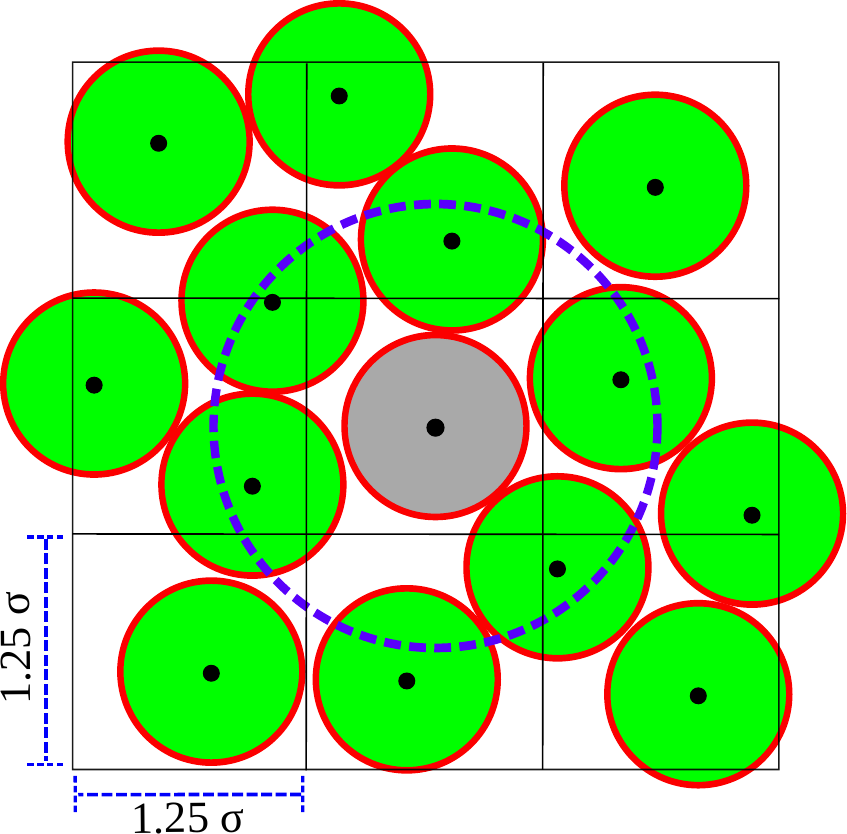}\\
\caption{\label{fig:density_grid} Schematic depiction of the 9 element density sub-grid used to compute the local density of particle $i$ (grey). All surrounding particles with coordinates (black dots) that lie within the sub-grid contribute to the local density, and hence to the $\epsilon$ value of particle $i$. The Lennard-Jones potential between the central particle $i$ and neighbours $j$ that lie beyond the cut-off region (blue circle), determined by the cut-off radius $r_c=1.2 \sigma$, is set to zero.}
\end{center}
\end{figure}
Our Brownian Dynamics simulations were performed using in-house software written in Java.
At each time-step of the density-dependent simulations, the interaction between any pair of particles, $i$ and $j$, depends on the local densities, $\rho_i$ and $\rho_j$, experienced by each particle. Local densities were computed on a $115 \sigma \times 115\sigma$ grid comprised of $92 \times 92$ grid elements each with a length of $1.25 \sigma$. The local density, $\rho_i$, experienced by particle $i$, was computed using a $3 \times 3$ element sub-grid centered around the element containing particle $i$ (Fig.\ref{fig:density_grid}). The value of $\rho_i$ was computed as the ratio of the sub-grid area, $9 \times (1.25 \sigma \times 1.25 \sigma$), to number of particles in the sub-grid, and the value of $\epsilon_i$ assigned to that particle was obtained via $\epsilon_i = \rho_i \epsilon'$. The geometric average, $\epsilon_{i,j} = \sqrt{\epsilon_i \epsilon_j}$, was then used to compute the Lennard-Jones potential (Eq. 1 of the main manuscript) for particle pairs separated by a distance less than the cut-off, $r_c=1.2\sigma$ (blue circle). The computation of neighbour interactions was made more efficient via implementation of a linked neighbour list \cite{Allen}.  

\section{\label{sec:nondimen_eqs}Non-dimensionalisation of the over-damped Langevin equation}
The position, ${\bf x}_i$ of an individual particle, $i$, at time $t$ was obtained in our simulations by solving the over-damped Langevin equation  
\begin{equation}
 \frac{{\rm d} {\bf x}_i}{{\rm d} t} = \beta D {\bf F}_i + \sqrt{2D} {\bf \eta}_i(t), 
\label{eq:langevin}
\end{equation}
where $D$ is the diffusion coefficient, $\beta = 1/k_BT$, ${\bf F}_i  = -\nabla \sum_{j \ne i}^N U_{ij}$ is the force on particle $i$ resulting from interactions with its $N-1$ neighbours, and ${\bf \eta}_i(t)$ is a unit variance white noise variable with $\langle \eta_i(t) \rangle=0$ and $\langle \eta_{i;\alpha}(t) \eta_{i;\beta}(t') \rangle = \delta_{\alpha, \beta}\delta(t-t')$ with $\alpha,\beta = x,y$. 

To integrate this equation numerically in our particle-based simulations, we used the discretised form of Eq. \ref{eq:langevin} for each component of the position vector ${\bf x_i}$  
\begin{equation}
\label{eq:discretisedBD}
 {\rm d} {x}_i = \beta D F_i \Delta t + \sqrt{2D} \tilde{\eta} \sqrt{\Delta t}, 
\end{equation}
where the dimensionless noise, $\tilde{\eta}(t)$, is a Gaussian random variable with mean $0$ and variance of $1$ \cite{grassia1995computer}. For computational efficiency, the values of the random variable were generated from a uniform distribution with range $[-\sqrt{3},\sqrt{3}]$ in order to ensure a mean value of $0$ and a variance of $1$ \cite{grassia1995computer}. 

We non-dimensionalised this equation by rescaling the variables $x$, $t$, and $F$ via the following transformations 
\begin{subequations}
\begin{align}
 x &\to \alpha \tilde{x} \\
 t &\to \tau \tilde{t} \\
 F &\to \phi \tilde{F}.
 \end{align}
\end{subequations}
Using $\sigma$, $\frac{\sigma^2}{D}$, and $k_BT$ as the basic units of length, time, and energy respectively, gives
\begin{subequations}
\begin{align}
 \alpha &=  \sigma \\
 \tau &= \frac{\sigma^2}{D} \\
 \phi &= \frac{k_BT}{\sigma}
 \end{align}
\end{subequations}
so that the rescaled variables are given by
\begin{subequations}
\begin{align}
 x &\to \sigma \tilde{x} \\
 t &\to  \frac{\sigma^2}{D} \tilde{t} \\
 F &\to  \frac{k_BT}{\sigma}\tilde{F}.
 \end{align}
 \label{rescaling1}
\end{subequations}
Substitution of Eqs. \ref{rescaling1} into Eq. \ref{eq:discretisedBD}, gives
\begin{equation}
\label{eq:sub_discretisedBD}
 {\rm d} \tilde{x}_i = \frac{1}{\sigma}\frac{D}{k_BT} \frac{k_BT}{\sigma} \frac{\sigma^2}{D} \tilde{F}_i \Delta \tilde{t} + \frac{1}{\sigma}\frac{\sqrt{\sigma^2}}{\sqrt{D}}\sqrt{2D} \tilde{\eta}_i(t) \sqrt{\Delta \tilde{t}}, 
\end{equation}
which is equivalent to the non-dimensionalised form
\begin{equation}
\label{eq:non_dim_discretisedBD}
 {\rm d} \tilde{x}_i = \tilde{F}_i \Delta \tilde{t} + \sqrt{2} \tilde{\eta}_i(t) \sqrt{\Delta \tilde{t}}. 
\end{equation} 

The use of $k_BT$ as the basic unit of energy rescales the interaction strength, $\epsilon$, in the Lennard-Jones potential to the dimensionless quantity $\tilde{\epsilon}$ via the transformation $\epsilon \to k_BT\tilde{\epsilon}$. The non-dimensionalised form of the Lennard Jones potential then has the form

\begin{equation}\label{eq:lj}
\tilde{U}(\tilde{r}) = 4\tilde{\epsilon}\left[(1/\tilde{r})^{12} - (1/\tilde{r})^{6} - \tilde{U}_{\rm c}\right],
\end{equation}
where, $\tilde{r}$ is the distance between the two particles, and $\tilde{U}_c$ is the cut and shifted term that ensures that the interaction energy at the cut-off distance, $r_c=1.2$, is zero. 
Considering $\epsilon$ as a function of the density, $\epsilon = \rho \times \epsilon'$, in our 2-dimensional density-dependent simulations, gives rise to the following rescaling transformations
\begin{subequations}
\begin{align}
\epsilon' & \to k_BT \sigma^2 \tilde{\epsilon} \\
\epsilon & \to  \frac{k_BT \sigma^2}{\sigma^2} \tilde{\epsilon}.
 \end{align}
\end{subequations}

\section{\label{sec:Equilibration}Equilibration}
\begin{figure*}[t!]
\begin{center}
\includegraphics[scale=1.0]{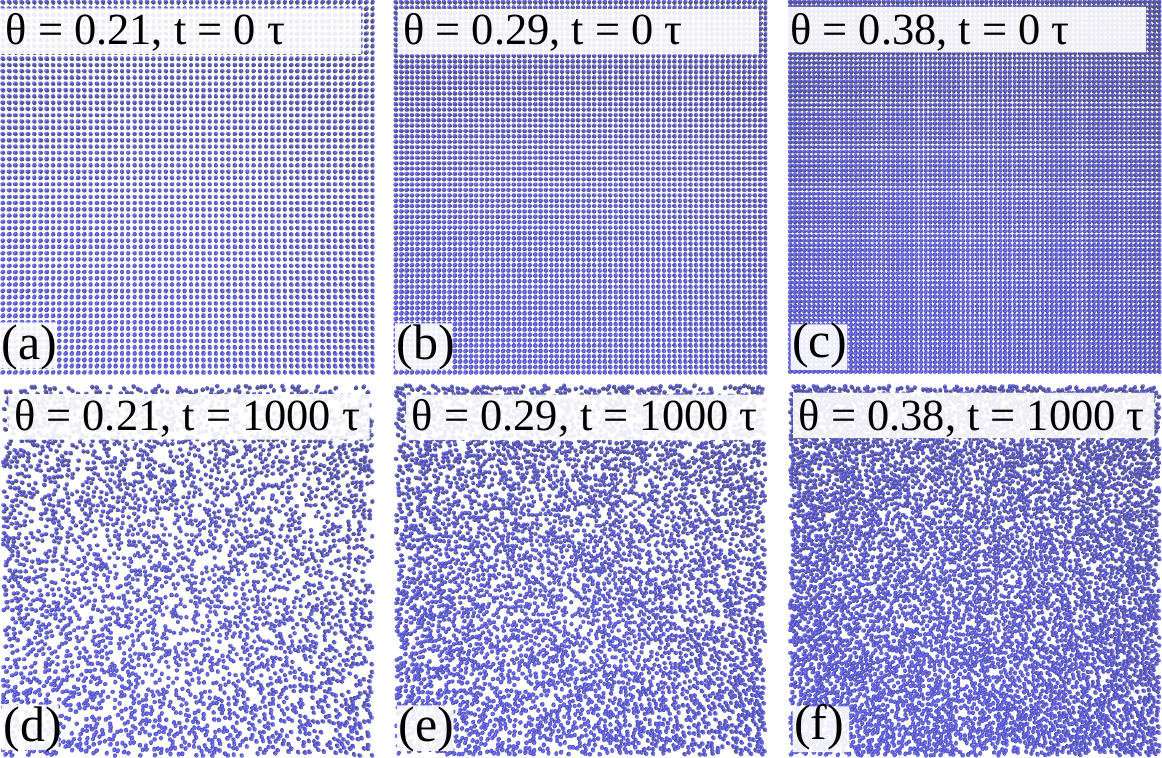}\\
\caption{\label{fig:lattice_confs} Relaxation of the systems from initial lattice configurations. (a)-(c) Initial lattice configurations for systems with area fraction $\theta =0.21$, $\theta =0.29$, and $\theta=0.38$. (d)-(f) Corresponding equilibrated configurations after ``equilibration'' runs of $1000\tau$ with WCA potential. This procedure was repeated 10 times for each area fraction. The resulting configurations were used to initialise 10 replicate simulations for each system under study.}
\end{center}
\end{figure*}

\begin{figure*}[t!]
\begin{center}
\includegraphics[scale=1.0]{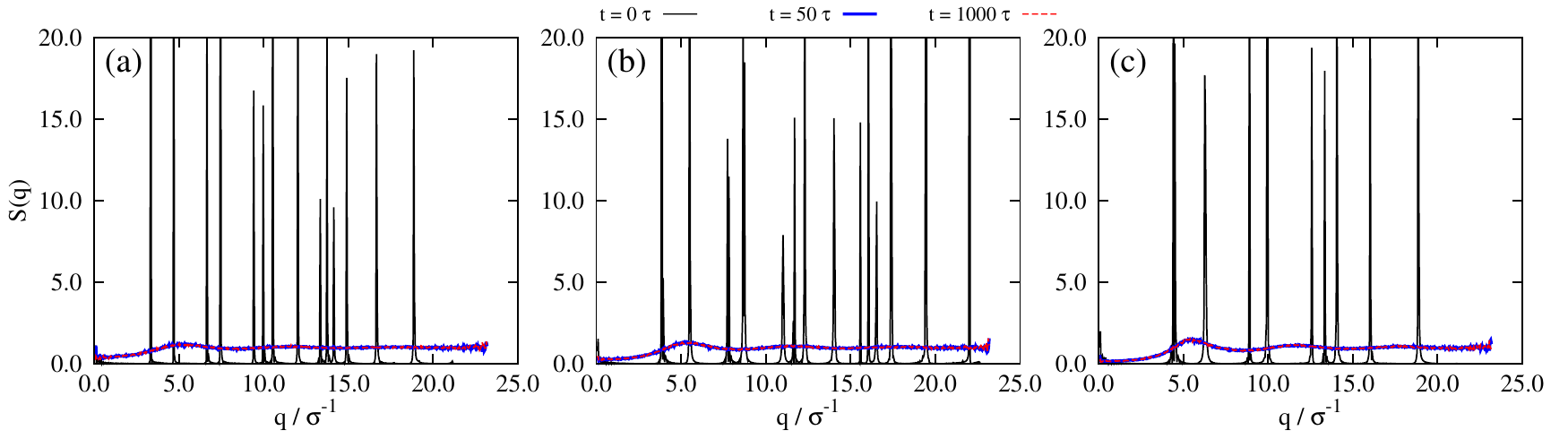}\\
\caption{\label{fig:relax_sks} Static structure factors, $S(q)$, corresponding to the relaxation of the systems from their initial lattice configurations (Figs. \ref{fig:lattice_confs}). (a) $\theta = 0.21$, (b) $\theta =0.29$, (c) $\theta=0.38$. An equilibration period of $1000 \tau$ is clearly sufficient for the initial structure imposed by the lattice configuration to relax. The different colours correspond to various time points of the simulation.}
\end{center}
\end{figure*}
The Brownian dynamics simulation method outlined above is stochastic due to the noise variable, $\eta$, and therefore many simulation runs should be performed in order to generate acceptable statistical data. We therefore performed 10 simulations for every parameter set $\theta$, $\epsilon'$ ($\epsilon$), in our density-dependent (-independent simulations). Stochasticity in each individual simulation was implemented by the use of a random-number generator, each of which was initialised with a different seed. 

To initiate our simulations of systems of particles interacting with density-dependent and -independent potentials, we first generated 10 equilibrated starting configurations for each of three area fractions. To do this, $N$ particles corresponding to area fractions $\theta=0.21 (N=3600)$, $\theta=0.29 (N=4900)$, and $\theta=0.38 (N=6400)$, were arranged on a 2D square lattice within a box of dimensions $115 \sigma \times 115\sigma$ (see Figs. \ref{fig:lattice_confs}(a)-(c)). These systems were then relaxed  for $1000 \tau$ using the Brownian Dynamics simulation method (time-step of $1 \times 10^{-4} \tau$) outlined in Section \ref{sec:nondimen_eqs} above and in the main manuscript. During these equilibration runs, particles interacted via the soft repulsive Weeks-Chandler-Andersen potential \cite{wca} 
\begin{equation}
   U(r) =
\begin{cases}
 4\epsilon \left[(\sigma/r)^{12} - (\sigma/r)^{6} \right]+ \epsilon& \text{if $r < 2^{\frac{1}{6}}\sigma$} \\
   0 & \text{if $r \ge 2^{\frac{1}{6}}\sigma$}, 
\end{cases}
\end{equation}
which when using $\frac{\epsilon} {k_BT} = \tilde{\epsilon} = 1$ gives the non-dimensionalised form 
\begin{equation}
   \tilde{U}(\tilde{r}) = 
\begin{cases}
 4\left[(1/{\tilde r})^{12} - (1/{\tilde r})^{6} \right]+ 1 & \text{if $\tilde{r} < 2^{\frac{1}{6}}$} \\
   0 & \text{if $\tilde {r} \ge 2^{\frac{1}{6}}$}.
\end{cases}
\end{equation}

Figures \ref{fig:lattice_confs}(d)-(f) show representative final configurations resulting from this relaxation procedure. It is clear from Figs. \ref{fig:relax_sks}, which show the static structure factor, $S(q)$ (see Sec. \ref{sec:structure_factors}), at three different time points during the simulations depicted in Figs. \ref{fig:lattice_confs}(a)-(c), that $1000 \tau$ is sufficient for the systems to relax from their initial lattice configurations. This relaxation procedure was performed 10 times for each area fraction, and the ``equilibrated'' final configurations were then used as starting configurations for simulations with the density-dependent and -independent Lennard-Jones interactions turned on. For example, to generate data for the density-dependent system $\theta=0.29$, $\epsilon'=40$, the 10 ``equilibrated'' configurations generated from the relaxation of the $\theta=0.29$ system, were used to initialised 10 ``production-run'' simulations, each with a different set of random numbers,  for this parameter set. 

\section{\label{sec:structure_factors} Structure factor computation}
Static structure factors were computed via 
\begin{equation}
 S(q) = N^{-1} \langle \rho(q) \rho(-q) \rangle,
\end{equation}
where $q$ is the magnitude of the wave vector ${\bf q} = (2\pi/L)(k_x,k_y)$, with $L=115\sigma$ being the box length, and $k_x$ and $k_y$ integers. The reciprocal space density, $\rho(q)$, is given by the spatial Fourier transform of the number density via 
\begin{equation}
 \rho(q) = \sum_{i=1}^N \exp(i{\bf q}\cdot {\bf r}_i).
\end{equation}

The structure factors at time $t=0$ in Figs. \ref{fig:relax_sks} were computed for the initial lattice configuration as described in Section \ref{sec:Equilibration} above.
The structure factors at each time $t>0$ were computed as the average of configurations sampled every $10\tau$ during time windows of length $50\tau$ in any one simulation. For example, the structure factors at $t=50 \tau$ were computed as the average over the $5$ simulation configurations sampled every $10\tau$ between $t=10 \tau \to 50\tau$. The structure factors at $t=1000 \tau$ were computed as the average over the $5$ simulation configurations sampled every $10\tau$ between $t=960 \tau \to 1000\tau$. 

For density-dependent simulations (see Sec. \ref{sec:all_vfs_DD}), in which a smaller time-step of $2.5\times 10^{-5} \tau$ is used, the structure factors (Figs. \ref{fig:sk_phase_diagram}) at each time $t>0$ were computed as the average of configurations sampled every $2.5\tau$ during time windows of length $60\tau$ in any one simulation, e.g., the $S(q)$ at $t=62.5 \tau$ was computed as the average over the $25$ simulation configurations sampled every $2.5\tau$ between $t=2.5 \tau \to 62.5\tau$; at $t=625 \tau$, $S(q)$ was computed as the average over the $25$ simulation configurations sampled every $2.5\tau$ between $t=565 \tau \to 625\tau$; and so on. 

\section{\label{sec:density}Simulation run lengths}
\begin{figure}[t!]
\begin{center}
\includegraphics[scale=1.0]{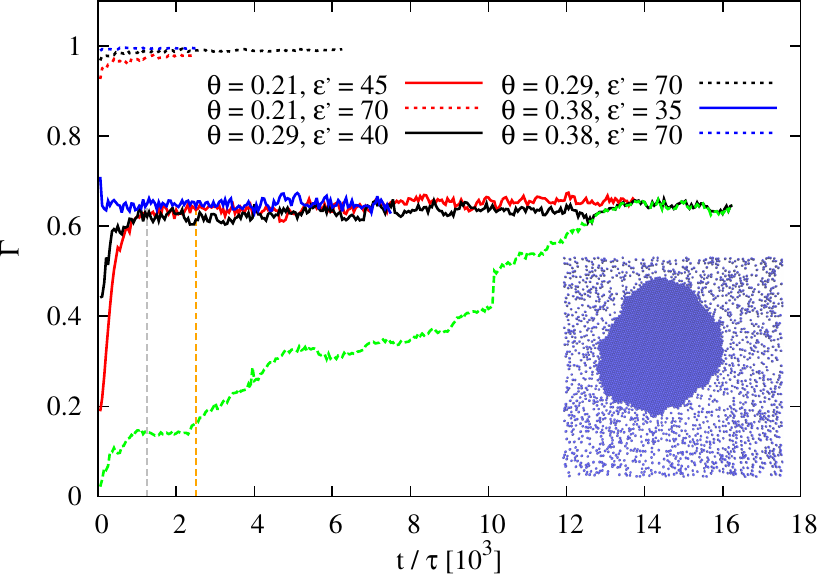}\\
\caption{\label{fig:frac_parts_clust} The fraction of particles in aggregates, $\Gamma$, as a function of time. $\Gamma$ approaches 1 for systems in the ``gel-like'' regime, $\epsilon'=70$. In the phase separating regimes, $\epsilon'=45,40,35$, for systems of area fraction $\theta=0.21,0.29,0.38$ respectively, $\eta$ has a steady-state value of $\sim0.65$. The time for the systems to reach this steady-state is $\sim 2500 \tau$ for the $\theta=0.21$ system (dashed orange curve), and $\sim 1250 \tau$ for the $\theta=0.29$ and $\theta=0.38$ systems (dashed grey curve). These times were considered long enough for simulation runs investigating the early aggregation behaviour in these systems. The dashed green curve shows the ratio of the maximum aggregate size to the number of particles, $N_a^{max}/N$, for the $\theta =0.29, \epsilon'=40$ system. This value equals $\Gamma$ when coarsening and coalescence processes result in one phase separated cluster in the system after $1.2 \times10^4 \tau$ (inset).}
\end{center}
\end{figure}

To explore the early phase separation behaviour in our density-dependent and density-independent simulations, we chose as our run-time, $t_{run}$, the approximate time required for the fraction of particles in clusters, $\Gamma$, to reach a steady state. Figure \ref{fig:frac_parts_clust} shows the time evolution of $\Gamma$ during density-dependent simulations for the systems in the ``gel-like'' ($\epsilon'=70$), and phase separating regimes ($\epsilon'=45$, $\epsilon'=40$, and $\epsilon'=35$) at $\theta=0.21$, $\theta=0.29$, and $\theta=38$.

The steady state value of $\Gamma$ is indicative of the extent of the phase separation. For example, small $\Gamma \to$ ``gas rich'' (data not shown); high $\Gamma \to$ ``gel-like''; and intermediate $\Gamma$ is indicative of a phase separated state consisting of aggregates immersed in a sea of particles. In this intermediate steady state regime, where $\Gamma \to 0.65$ (the main focus of this study), one can see that $\Gamma$ has reached a steady state for $\theta = 0.21$ by $\sim 2500 \tau$, whereas in the systems of higher area fractions, $\theta = 0.29$ and $\theta = 0.38$, equilibration of $\Gamma$ is attained at the earlier time of $\sim 1250\tau$. Beyond this time, $\Gamma$ has reached a steady state and coarsening and coalescence changes the distribution of aggregate sizes until there is a single phase separated domain by $12000\tau$ (inset). 

Although, these growth processes lead to an increase in the maximum aggregate size $N_a^{max}$ (the green dashed curve shows $\frac{N_a^{max}}{N}$ for $\theta=0.29, \epsilon' =40$), the fraction of particles in clusters remains relatively constant. 
The time evolution of $\frac{N_a^{max}}{N}$ (green dashed curve) provides greater insight into this redistribution process. At times greater than $12000 \tau$, this ratio becomes equal to $\Gamma$ meaning that there is one large cluster in the system (inset). Jumps in the green curve are indicative of coalescence events whereas a steady increase points to coarsening.

\section{\label{sec:all_vfs_DD}Phase separation behaviour and structure emerging from density-dependent simulations}
\begin{figure*}[t!]
\begin{center}
\includegraphics[scale=0.9]{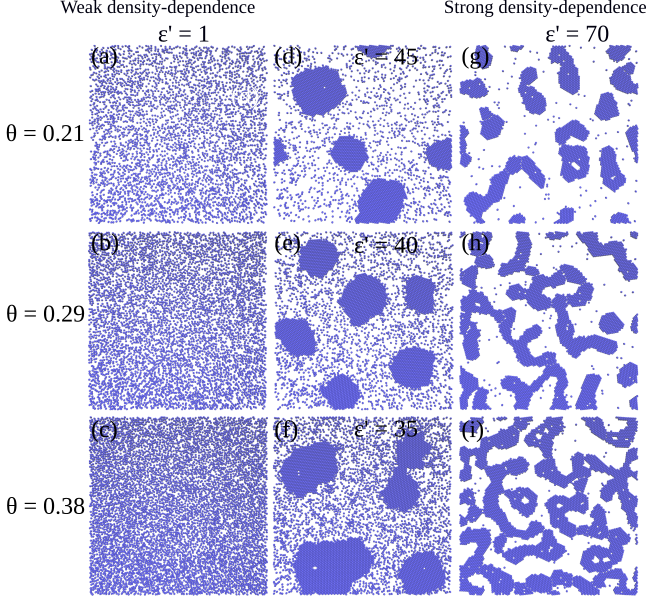}\\
\caption{\label{fig:full_phase_diagram} Simulation snapshots of systems of Brownian particles interacting via density-dependent potential. Snapshots are representative of simulation configurations at times $t_{run}=2500 \tau$ ($\theta=0.21$) and $t_{run}=1250 \tau$ ($\theta=0.29$ and $\theta=0.38$).}
\end{center}
\end{figure*}
\begin{figure*}[t!]
\begin{center}
\includegraphics[scale=1.0]{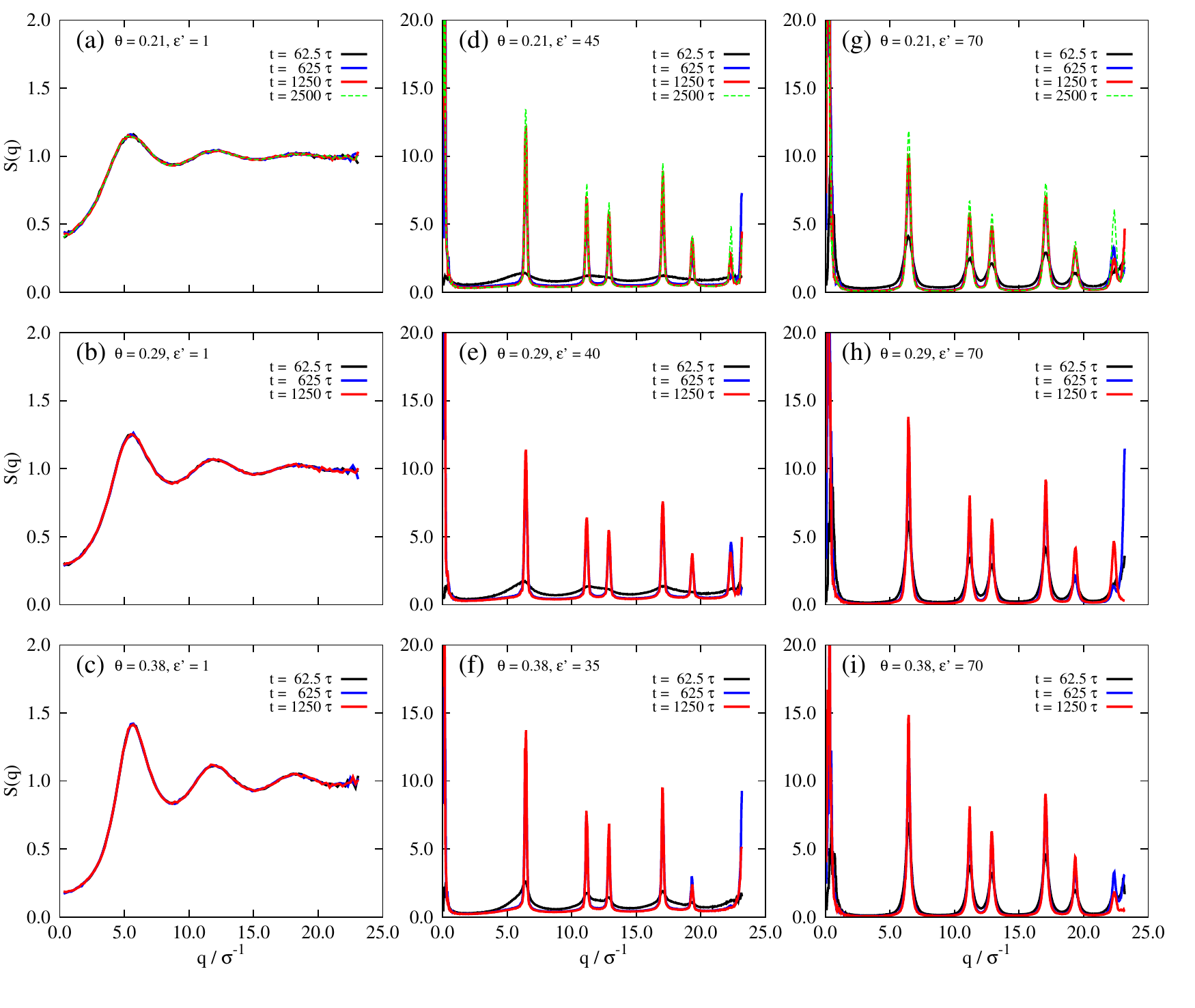}\\
\caption{\label{fig:sk_phase_diagram} Static structure factors, $S(q)$, corresponding to various time points during the evolution of the density-dependent simulation towards the states depicted in Fig. \ref{fig:full_phase_diagram}}.
\end{center}
\end{figure*}

Figure \ref{fig:full_phase_diagram} shows configurations at $t_{run}$ for the three area fractions $\theta=0.21$ ($t_{run}=2500\tau$), $\theta=0.29$ ($t_{run}=1250\tau$), and $\theta=0.38$ ($t_{run}=1250\tau$) for values of $\epsilon'$ corresponding to: the ``gas-phase'' regime ($\epsilon'=1$); the ``phase-separation'' regime ($\epsilon'=45,40,35$); and the ``gel-like'' regime ($\epsilon'=70$). 
At $\epsilon' = 1$, the three systems are in a ``gas-like'' phase with ordering, characterised by the structure factor $S(q)$, consistent with that of a simple colloidal dispersion or hard sphere fluid (Figs. \ref{fig:sk_phase_diagram} (a)-(c)). At $\epsilon' = 70$, the aggregates are more elongated and tend to ``gel-like'' states with increasing $\theta$. Well-defined structure in these condensed phases is reflected by the peaks in $S(q)$ (Figs. \ref{fig:sk_phase_diagram} (g)-(i)). 

At intermediate values of $\epsilon '$ (Figs. \ref{fig:full_phase_diagram} (d)-(f)), it is  clear that the systems are undergoing phase separation into condensed and non-condensed phases, with the emergence of the former being determined by the presence of well-defined peaks in $S(q)$ (Figs. \ref{fig:sk_phase_diagram} (d)-(f)). 
Strikingly, the particles are distributed such that condensed phase aggregates are immersed in a dilute ``gas''. Phase separation is initialised in the density-dependent systems with the widespread formation of aggregates that emerge with no apparent barrier to formation (see Supplemental Material \cite{SI} for movie). Such behaviour is consistent with spinodal decomposition but rather than forming elongated structures typical of this process, the aggregates bear more of a resemblance to those that arise from nucleation (see Supplemental Material \cite{SI} for movie).

\section{\label{sec:all_vfs_DI} Density independent simulations}
\begin{figure}[t!]
\begin{center}
\includegraphics[scale=1]{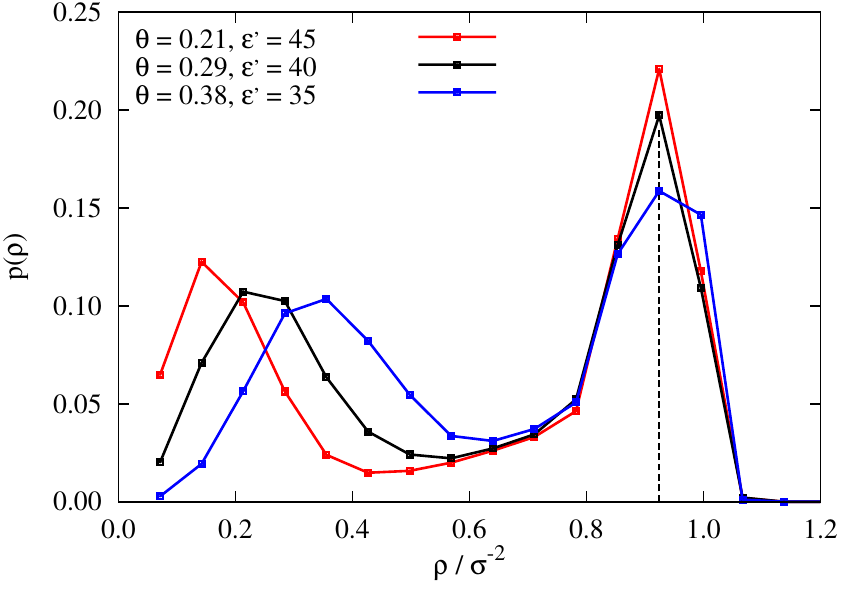}\\
\caption{\label{fig:dens_hists} Normalised probability distributions of the local densities sampled in the 10 final configurations at $2500 \tau$ ($\theta=0.21$) and  $1250 \tau$ ($\theta=0.29,0.38$). The dashed vertical line corresponding to $\rho=0.924$, gives rise to $\epsilon=41.58, 36.96, 32.34$ for $\theta=0.21,0.29,0.38$ with which to run density-independent simulations.}
\end{center}
\end{figure}

In any given density-dependent simulation, a range of $\epsilon$ values is sampled by the interacting particles. Therefore, to compare our results to those for systems of particles interacting independently of the density, we carried out density-independent simulations with Lennard-Jones interaction strength $\epsilon$ corresponding to the values of $\epsilon$ sampled (via $\epsilon = \rho ({\bf x}) \epsilon'$) in the final configurations of the phase separating regimes of our density-dependent simulations (Figs. \ref{fig:full_phase_diagram}(d-f)). 

Focusing here on the system with $\theta=0.29$ and $\epsilon'=40$, the procedure was performed as follows. From the 10 final configurations at $t_{run}=1250 \tau$ (one of which is shown in Fig. \ref{fig:full_phase_diagram} (e)), a normalised distribution, $p(\rho)$, of the local density experienced by each particle (Fig. \ref{fig:dens_hists}) was constructed. For the 15 values of $\rho$ sampled in these final configurations, a value of $\epsilon$ was generated for each via $\epsilon = \rho ({\bf x}) \epsilon'$ (see Table \ref{tab:epsilon_values}). For each value of $\epsilon$, 10 density-independent simulations were performed. This procedure was repeated for density-independent simulations corresponding to the $\theta=0.21$, $\epsilon'=45$, and the $\theta=0.38$, $\epsilon'=35$ systems. As will be discussed in more detail below, the value of $p(\rho)$ at each value of $\rho$ was also used as the weight for the construction of a weighted linear superposition of density-independent aggregate size distributions (see Fig. 3(c) of main manuscript).  

Figure \ref{fig:dens_hists} shows $p(\rho)$ from the density-dependent systems taken at time $t_{run}$. The two peaks in $p(\rho$) correspond to the phase-separated states shown in Figs. \ref{fig:full_phase_diagram}(d)-(f). Although increasing $\theta$ has little effect on the location of the high density peak due to limitations in packing, it does have the effect of increasing the density of the ``non-condensed'' phase. Increasing $\theta$ also leads to a decrease in the bimodal nature of the distributions because more particles are available to sample intermediate values of $\rho$.

\begin{table*}[h!]
\begin{center}
\begin{tabular}{l l |l |l |l |l |l |l |l |l }
& &\multicolumn{2}{c}{$\theta = 0.21$} & &\multicolumn{2}{c}{$\theta=0.29$} & & \multicolumn{2}{c}{$\theta=0.38$} \\ \hline
  $\rho$  & &   $\epsilon = \epsilon' \times \rho$   &   $w$   & &   $\epsilon = \epsilon' \times \rho$   &   $w$    & &   $\epsilon = \epsilon' \times \rho$   &   $w$   \\ \hline

0.071 &\hspace{0.25cm} & 3.2 & 0.065 & \hspace{0.25cm} & 2.84 & 0.020 &\hspace{0.25cm} & 2.49 & 0.003 \\
0.142 & &6.40 & 0.123 & & 5.68 & 0.071 & & 4.97 & 0.019\\
0.213 & &9.59 & 0.102 & & 8.52 & 0.107 & & 7.46 & 0.056\\
0.284 & &12.78 & 0.056 & & 11.36 & 0.102 & & 9.94 & 0.096\\
0.356 & &16.02 & 0.024 & & 14.24 & 0.064 & & 12.46 & 0.104\\
0.427 & &19.22 & 0.015 & & 17.08 & 0.036 & & 14.95 & 0.082\\
0.498 & &22.41 & 0.016 & & 19.92 & 0.024 & & 17.43 & 0.055\\
0.569 & &25.61 & 0.020 & & 22.76 & 0.021 & & 19.92 & 0.034 \\
0.640 & &28.8 & 0.026 & & 25.60 & 0.027 & & 22.40& 0.031\\
0.711 & &32.00 & 0.033 & & 28.44 & 0.034 & & 24.89 & 0.037\\
0.782 & &35.19 & 0.046 & & 31.28 & 0.052 & & 27.37 & 0.051\\
0.853 & &38.39 & 0.134 & & 34.12 & 0.131 & & 29.86 & 0.126\\
0.924 & &41.58 & 0.221 & & 36.96 & 0.197 & & 32.34 & 0.159\\
0.996 & &44.82 & 0.118 & & 39.84 & 0.109 & & 34.86 & 0.146\\
1.067 & &48.02 & 0.002 & & 42.68 & 0.002 & &  37.34& 0.001\\
\hline
\end{tabular}
\caption[]{Parameters derived from density-dependent simulations to be used for input (analysis) to (of) density-independent simulations.}
\label{tab:epsilon_values}
 \end{center}
 \end{table*}

\begin{figure*}[t!]
\begin{center}
\includegraphics[scale=1.0]{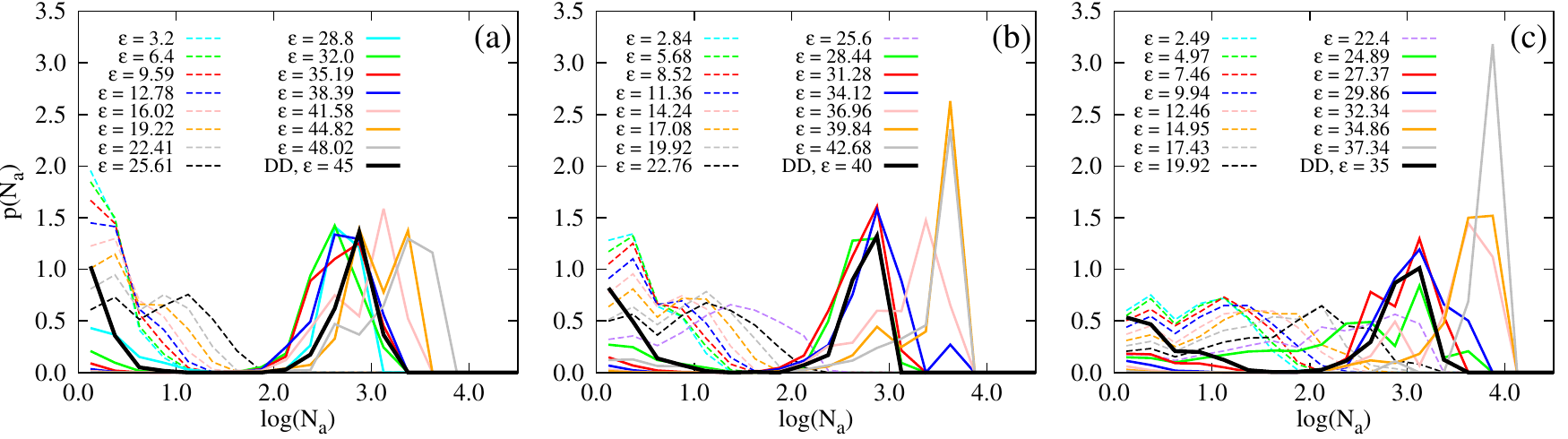}\\
\caption{\label{fig:cons_agg_dist} Distribution of aggregate sizes from density-independent simulations at $2500 \tau$ for $\theta=0.21$ (a) and $1250 \tau$ for $\theta =0.29$ (b) and $\theta=0.38$ (c). Dashed curves correspond to ``non-aggregating'' regimes discussed in the text. Each distribution was generated from the final configurations of 10 replicate simulations. The aggregate peak of the purple dashed curve at $\sim {\rm log}(3.0)$ in (c) is a result of aggregates present in 4 of the 10 simulations at $\epsilon=22.4$. Aggregation in this case resulted from nucleation events thus only occurring in a few of the simulations.}
\end{center}
\end{figure*}

Figure \ref{fig:cons_agg_dist} shows the aggregate size distributions resulting from density-independent simulations at $t_{run}=1250\tau$ ($\theta=0.29,0.38$) and $2500 \tau$ ($\theta=0.21$). The solid lines represent phase separated regimes in which condensed phase aggregates or gel-like networks are clearly present. The distributions shown with dashed lines, the ``non-aggregating regime'',  fall into three categories: 1) No permanent condensed phase, peaks located at $\log(A)<1$; 2) Transient aggregates, peaks located between $1<\log(A)<\sim 2.5$: here the ``seeds'' of aggregates dissolve upon formation and therefore fail to initialise; and 3) Condensed phase aggregates present in some but not all of the simulations, e.g., at $\theta=0.29$, $\epsilon=22.4$ (dashed purple curve in (c)), the peak at $\sim {\rm log}(3.0)$ is a result of aggregates that have formed via nucleation in $4/10$ simulations. At all three area fractions, the size of the ``aggregates'' increases with increasing $\epsilon$. It is also evident that the lowest value of $\epsilon$ needed to induce phase separation decreases with increased area fraction $\theta$. In the ``non-aggregating'' regimes, the increased sampling of transient aggregate sizes with increasing $\theta$ gives rise to flatter distributions. It is evident that the density-independent simulations fail to produce the distributions that emerge from the density-dependent simulations (solid black curves), i.e., two peaks corresponding to condensed and non-condensed phases with little sampling of intermediate aggregate sizes.     

\begin{figure*}[t!]
\begin{center}
\includegraphics[scale=0.9]{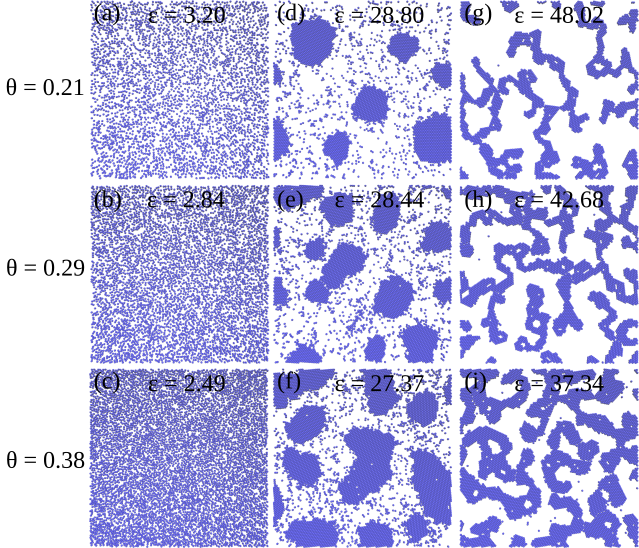}\\
\caption{\label{fig:const_phase_diagram} Simulation snapshots of systems of Brownian particles interacting via standard cut and shifted Lennard-Jones potential. Snapshots are representative of simulation configurations at times $t_{run}=2500 \tau$ ($\theta=0.21$) and $t_{run}=1250 \tau$ ($\theta=0.29,0.38$) for ``non-aggregating'' (first column), ``phase separating'' (middle column), and ``gel-like'' regimes.}
\end{center}
\end{figure*}

Figure \ref{fig:const_phase_diagram} shows a representative sample of final configurations resulting from density-independent simulations in the: ``non-aggregating'' (first column), ``aggregating'' (middle column), and ``gel-like'' (third column) regimes. Configurations in the aggregating regimes (middle column) are representative of those distributions in Figs. \ref{fig:cons_agg_dist} that most closely resemble the distributions generated from the corresponding density-dependent simulations (solid black curves in Figs. \ref{fig:cons_agg_dist}). Comparing configurations from the density-dependent and density-independent simulations, (Figs. \ref{fig:full_phase_diagram}(d)-(f) and Figs. \ref{fig:const_phase_diagram}(d)-(f)), we see that density-dependent cohesion provides a means by which condensed phase aggregates can coexist with greater numbers of ``single particles'', i.e., a richer non-condensed phase, evident by the higher value of the ``single particle'' peak of the black solid curves in Figs. \ref{fig:cons_agg_dist}.

To assess whether the bimodal distributions generated in our density-dependent simulations $\theta=0.21,\epsilon'=45$; $\theta=0.29,\epsilon'=40$; and $\theta=0.38,\epsilon'=35$ can be viewed as just a simple combination all of the individual distributions in each of Figs. \ref{fig:cons_agg_dist} (a)-(c), we constructed weighted linear superpositions of these distributions, $p_{ls}$ (Fig. \ref{fig:lin_sup_dists}), via 
\begin{equation}
 p_{ls}(N_a) = \sum_{i=1}^{15} p_{\epsilon_i}(N_a) \times w_{\epsilon_i}, 
\end{equation}
where $p_{\epsilon_i}$ is the aggregate size distribution for a given value of $\epsilon$, and $w_{\epsilon_i}$ is the corresponding weight of that value (Table \ref{tab:epsilon_values}) given by the probabilities in Fig. \ref{fig:dens_hists}, e.g., for the system with $\theta=0.29$ and $\epsilon'=40$, the linear superposition distribution at $N_a$ is given by 
\begin{equation}
 p_{ls}(N_a) = p_{\epsilon_1=3.2}(N_a) \times 0.065 + ... +   p_{\epsilon_{15}=42.68}(N_a) \times 0.002.
\end{equation}

Comparing the resulting linear superpositions with the distributions from the corresponding density-dependent simulations (Fig. {\ref{fig:lin_sup_dists}, solid and dashed curves respectively), it is clear that the aggregation behaviour emerging as a result of density-dependent cohesion is not a simple combination of the various $\epsilon$ values explored. 
Compared to the density-dependent simulation results, these superpositions show: a much broader range of aggregate sizes; a much more uniform sampling of aggregate sizes; and a greater sampling of larger aggregates.

\begin{figure}[t!]
\begin{center}
\includegraphics[scale=1]{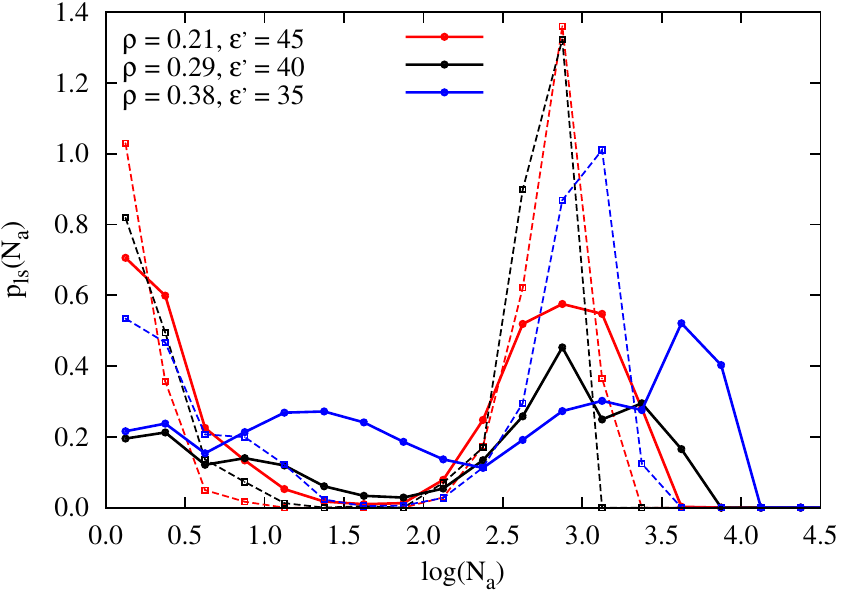}\\
\caption{\label{fig:lin_sup_dists} Aggregate size distributions generated from weighted linear superposition of density-independent simulations (solid curves) corresponding to the phase separating regimes of the density-dependent systems ($\theta=0.21, \epsilon'=45$, $\theta=0.29, \epsilon'=40$, $\theta=0.38, \epsilon'=35$). The original density-dependent aggregate size distributions are shown as dashed curves of the same colour.}
\end{center}
\end{figure}

\section{\label{sec:all_vfs} Landau Free Energies and Continuum Model}
\begin{figure*}[t!]
\begin{center}
\includegraphics[scale=1]{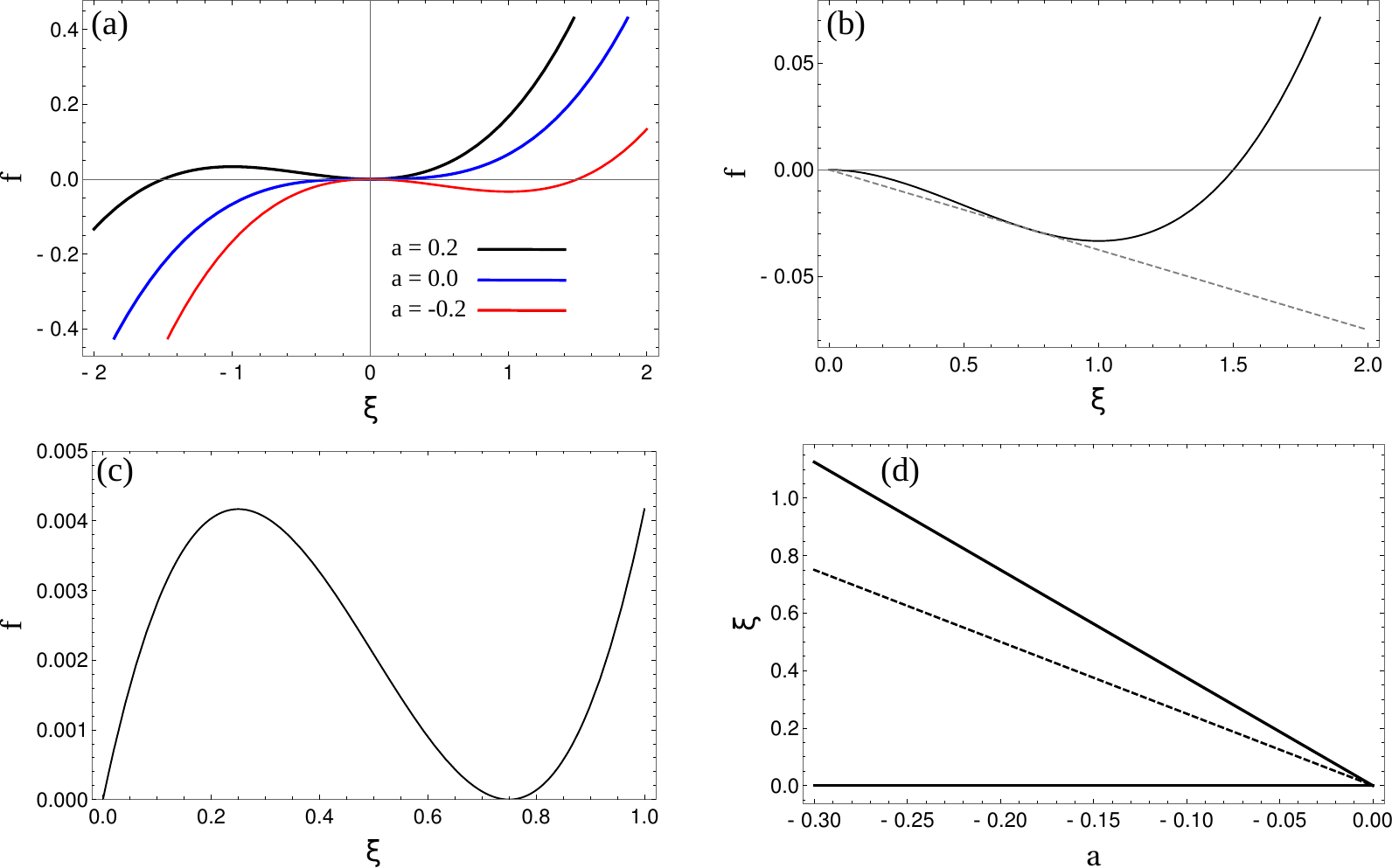}\\
\caption{\label{fig:dens_indep_landau} Free energy curves and phase diagram for density-independent Landau system. (a) Free energy density, $f(\xi)$, at three different values of the critical parameter $a$. (b) $f(\xi)$ and its common tangent line (dashed curve) for $a=-0.2$. (c) Tilted $f(\xi)$ computed via $f(\xi)-\mu_{\xi}\xi$. (d) Phase diagram resulting from the construction of binodal (solid) and spinodal lines (dashed). A value of $b=0.2$ was used for all data shown. All variables and parameters are in simulation units.}
\end{center}
\end{figure*}

To ascertain the parameters with which to perform our continuum simulations, we constructed phase diagrams by generating the binodal and spinodal lines in the plane of the critical parameters $a$ for the density-dependent case (and $b'$ for density-dependence), and the global density $\xi_0$. For both density-dependent and -independent systems, the binodal and spinodal lines were generated via examination of the homogeneous part of the Landau free-energy density. 

The homogeneous part of the free energy density for the density-independent system is  
\begin{equation}\label{eq:non_dens_dep}
f(\xi) =\frac{a}{2}\xi^2+\frac{b}{3}\xi^3.
\end{equation}
The order parameter, $\xi$, is a measure of the local density which, to be physically realistic, must satisfy the condition that $\xi \ge 0$.
The critical parameter, $a=\psi-\upsilon$ can be positive or negative depending on the interplay between entropic ($\psi$) and attractive ($\upsilon$) parts of the interaction, and governs the transition from disordered to ordered states.  The parameter $b$ is a positive constant. When coupled with the condition that $\xi \ge 0$, this ensures that the cubic term in Eq. \ref{eq:non_dens_dep} is the lowest order term required to stabilise the free energy. This is evident from Fig. \ref{fig:dens_indep_landau} (a), which shows how the functional form of the free energy density (Eq. \ref{eq:non_dens_dep}) changes with the critical parameter $a$. For $a>0$, the minimum of the free energy density is located at $\xi=0$, whereas for $a<0$, the minimum occurs at a positive finite value of $\xi$. 

In our system, the overall density $\xi_0$ is conserved according to $\int \xi {\rm d}{\bf x} = \xi_0 A$, where $A$ is total area. Therefore, the system cannot simply adjust its value of $\xi$ in order to minimise the free energy. However, for values of $\xi_0$ that correspond to regions of negative curvature in the free energy, the system \emph{can} lower its free energy by phase separating into coexisting condensed and non-condensed phases with local densities of $\xi_2$ and $\xi_1$ respectively, the values of which satisfy the common tangent construction (Fig. \ref{fig:dens_indep_landau} (b)). Subtraction of the common tangent line from the free energy has the effect of tilting the free energy curve so that the points of common tangent are the minima of the tilted free energy (Fig. \ref{fig:dens_indep_landau} (c)). 

Mathematically, the common tangent is constructed by equating both the chemical potential, $\mu$, and the pressure, $P$, of the two phases
\begin{align}\label{eq:com_tan}
\mu_{\xi_1} & = \mu_{\xi_2},\nonumber \\
 P_{\xi_1} & = P_{\xi_2}, 
\end{align}
where 
\begin{equation}
 \mu(\xi) = \frac{{\rm d}f(\xi)}{{\rm d} \xi}, 
\end{equation}
and 
\begin{equation}
 -P = f(\xi) - \mu(\xi)\xi.
\end{equation}
The solutions of Eqs. \ref{eq:com_tan} for all real positive values of $\xi$ give rise to a locus of points that form the binodal (coexistence) line. Given that there is only one stable minimum for the density-independent free energy (Fig. \ref{fig:dens_indep_landau} (a)), the condition $\xi>=0$ can be imposed
by setting $\xi_1=0$ for the noncondensed phase. Then to generate the condensed phase part of the bimodal line, we only need to solve the expression
\begin{equation}
 -P_{\xi_2} = f_{\xi_2} - \mu_{\xi_2}\xi = 0, 
\end{equation}
which has the solution $\xi_2 = -\frac{3a}{4b}$ (see solid lines in Fig. \ref{fig:dens_indep_landau} (d)). The spinodal line was constructed by finding the locus of inflection points of the free energy density according to  
\begin{equation}\label{eq:spinodal}
 \frac{{\rm d}^2f(\xi)}{{\rm d}\xi^2} = 0, 
\end{equation}
which gives the solution $\xi = -\frac{2a}{3b}$ (dashed line in  Fig. \ref{fig:dens_indep_landau} (d)). In these calculations, the value of $b$ was chosen, somewhat arbitrarily, to be $0.2$, so that the maximum value of the local density $\sim$ 1. Our density-independent continuum simulations were initialised from a homogeneous system using parameters $a=-0.2$ and $\xi_0=0.2$. This system lies in the phase separation region of the phase diagram in Fig. \ref{fig:dens_indep_landau} (d).  

\begin{figure*}[t!]
\begin{center}
\includegraphics[scale=1]{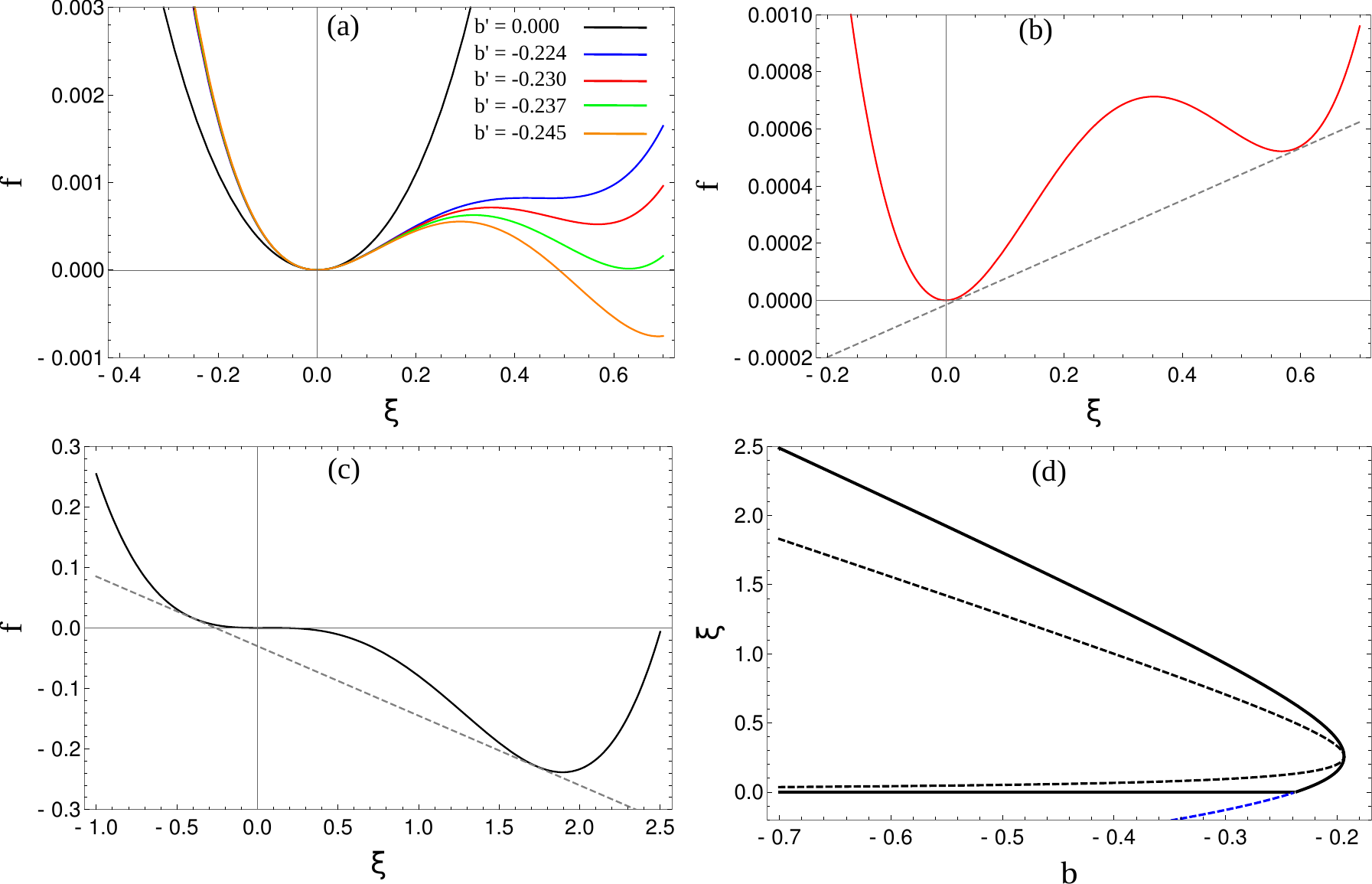}\\
\caption{\label{fig:dens_dep_landau} Free energy curves and phase diagram for density-dependent Landau system. (a) Free energy density, $f(\xi)$, at three different values of the critical parameter $b'$. (b) $f(\xi)$ and its common tangent line (dashed curve) for $b'=-0.23$. (c) $f(\xi)$ and its common tangent line (dashed curve) for $b'=-0.50$. (d) Phase diagram resulting from construction of binodal (solid) and spinodal (dashed) lines. The dashed blue line highlights negative values of $\xi$ resulting from common tangent construction. Values of $a'=0.05$ and $c=0.25$ were used for all data shown. All variables and parameters are in simulation units.}
\end{center}
\end{figure*}

We follow a similar protocol for constructing the phase diagram for the density-dependent system. The homogeneous part of the Landau free energy density for this system is given by (see main manuscript)
\begin{equation}\label{eq:dens_dep}
f=\frac{a'}{2}\xi^2+\frac{b'}{3}\xi^3 + \frac{c}{4}\xi^4,
\end{equation}
where $a'=\psi>0$, and $b' = b-\frac{3}{2}\upsilon'$ is the critical parameter that governs the phase transition. Given that the critical parameter is the coefficient of the cubic term, a quartic term with positive coefficient, $c>0$, must be present to stabilise the free energy. 
For values of $b'$ less than a critical value $b'_c \sim -0.194$ (data not shown), there are two stable minima in the density-dependent free energy density separated by a free energy barrier (Fig. \ref{fig:dens_dep_landau} (a)). One can construct common tangents for the free energy density via Eqs \ref{eq:com_tan}. Figures 
\ref{fig:dens_dep_landau} (b) and (c) show representative common tangent constructions that give rise to both positive and negative values of $\xi_1$ respectively. The latter (blue dashed line Fig. \ref{fig:dens_dep_landau} (d)) violates the condition $\xi>=0$, therefore in such instances, the value of $\xi_1$ is set to zero. The spinodal was again generated by solving Eq. \ref{eq:spinodal} and the resulting phase diagram is depicted in Fig. \ref{fig:dens_dep_landau} (d). In these calculations, the values of $a'$ and $c$ were chosen, somewhat arbitrarily, to be $0.05$ and  $0.25$ respectively so that the maximum value of the local density, $\xi \sim 1$. Our density-dependent continuum simulations were initialised from a homogeneous configuration with parameters $b'=-0.2$ and $\xi_0=0.3$. This lies in the phase separation region of the phase diagram in Fig. \ref{fig:dens_dep_landau} (d). 

In our simulations, the dynamics of $\xi$ were modelled using the Cahn-Hilliard equation \cite{Cahn-Hilliard}
\begin{equation}\label{eq:cahn_hilliard}
\frac{\partial \xi({\bf x},t)}{\partial t} =  M \nabla^2  \left(\frac{\delta }{\delta \xi ({\bf x})}\int_A f(\xi({\bf x})) {\rm d}{\bf x} \right)+ \nabla \cdot{\bf J}_r,
\end{equation}
where $M$ is the mobility, the term in brackets is the chemical potential, and ${\bf J}_r$ is a random flux which is spatially and temporally uncorrelated, with mean $\langle {\bf J}_r({\bf x},t) \rangle=0$, and variance $\langle {\bf J}_{r;\alpha}({\bf x},t) \cdot {\bf J}_{r;\beta}({\bf x},t') \rangle = \Lambda \xi \delta_{\alpha,\beta} \delta({\bf x}-{\bf x}')\delta(t-t')$, with $\alpha,\beta = x,y$. For simplicity, both $M=0.01$ and the random noise strength $\Lambda=0.1$ are kept constant. 

Equation \ref{eq:cahn_hilliard} was solved on an $L \times L$ grid ($128\le L \le 1024$) using standard finite difference simulations, with periodic boundary conditions. To map our continuum simulation units to real units we arbitrarily assign length-scales and time-scales to our simulation units such that we can compute an effective diffusion coefficient, $D_{eff}$, with which to compare to the passive (non-motile) diffusion coefficient, $D_p=0.2 \mu m^2 s^{-1}$, of the bacterium {\it E. coli} \cite{berg2008coli}. This species of bacteria has a length of $\sim 2 \mu m$ and divides approximately every $20$ minutes depending on the growth conditions \cite{Donachie}. 

We used a grid spacing, $l$, of $0.25$ simulation length units (SLU), which we can assigned to be equal to 
$1 \mu m$, thus giving a characteristic length scale, $l_c$, of $4 \mu m$. Our analysis was performed after $2\times 10^5$ simulation steps using a time-step, $\Delta t$, of $0.1$ simulation time units (STU) to give run-times, $t_{run}$, of $2\times 10^4$ STU. We assigned this time to be equal to $20$ minutes ($1.2\times10^3s$) in order to work within experimental time-scales, in which the number of cells is conserved (no growth and division). This gives rise to a characteristic time-scale, $t_c$, of $\frac{1.2\times10^3}{2\times 10^4}=6\times 10^{-2}$ seconds. We can use these characteristic time- and length- scales to compute the effective diffusion coefficient via $D_{eff} = \frac{l_c^2}{t_c}D_{s}$, where $D_s = Ma$ (simulation units of energy times inverse time). Using $M=0.01$ and $a=0.05$ (for the density-independent Landau free energy density), the value of $D_{eff}$ for our continuum simulations is $0.13 \mu {\rm m}^2 {\rm s}^{-1}$ which is comparable to that of {\it E. Coli}.

\bibliography{supplemental_material}